\documentclass[useAMS,usenatbib,usegraphicx]{mn2e}
\usepackage{epsfig}

\title[High-precision K-band photometry of HD209458]{High-precision K-band photometry of the secondary eclipse of HD209458}

\author[I.A.G. Snellen]{I.A.G. Snellen$^{1,2}$\thanks{E-mail: snellen@strw.leidenuniv.nl}\\
$^{1}$ Leiden Observatory, Postbus 9513, NL-2300RA Leiden, The Netherlands\\
$^{2}$ Institute for Astronomy, University of Edinburgh, Blackford Hill, Edinburgh, EH9 3HJ, UK\\}

\date{}

\begin{document}
\maketitle

\label{firstpage}

\begin{abstract}
Recently, mid-infrared Spitzer observations have been presented that show the 
light decrement due to the passage of a planet behind its host star.
These measurements of HD209458b and TrES-1 are the first detections of 
direct light from an extra-solar planet. Interpretation of these results in
terms of planet equipartition temperature and bond albedo is however 
strongly model dependent and require additional observations at shorter 
wavelengths. Here we report on two attempts to detect the 
secondary eclipse of HD209458b from the ground in K-band, using the 
UK InfraRed Telescope (UKIRT). A photometry precision of 0.12\%  
relative to two nearby reference stars  was reached during both occasions, 
but no firm detection of the eclipses were obtained. The first observation 
shows a flux decrement of $-$0.13$\pm$0.18\%, and the second of 
$-$0.10$\pm$0.10\%. A detailed description of the observing strategy, 
data reduction and analysis is given, and a discussion on how the 
precision in ground-based K-band photometry could be further improved. 
In addition we show that the 
relative photometry between the target and the reference stars between the 
two epochs is consistent down to the $<$0.1\% level, which is 
interesting in the light of possible near-infrared surveys to search 
for transiting planets around M and L dwarfs.
\end{abstract}


\section{Introduction}

The large majority of the extra-solar planets known to date have been 
discovered using the radial velocity technique, providing a wealth
of information on the orbital properties of the exoplanet population
(e.g. Marcy et al. 2005). 
However, so far only those planets that {\sl transit} their host star 
have given also an insight in the properties of the planets themselves. 
Firstly, the observations of the transit allows the determination of the 
inclination of the orbit, and subsequently the planetary mass, radius, and 
density. Furthermore, precise determination of the transit depth as function of
wavelength can reveal atmospheric constituencies. 
For the brightest transiting exoplanet system known so far, HD209458b, 
this has resulted in the detection 
of Sodium in the exoplanet's atmosphere (Charbonneau et al. 2002), 
and hydrogen, carbon, and oxygen in its 
evaporating exosphere (Vidal-Madjar et al. 2003; 2004)

Recently, observations of transiting exoplanet systems have resulted in a new 
breakthrough in exoplanet research. By observing secondary eclipses of 
the two brightest known transiting exoplanets, HD209458 and TrES-1,
the light contribution from the planets have been determined. 
While Charbonneau et al. (2005) observed a secondary eclipse of TrES-I
with Spitzer at 4.5 and 8 micron, Deming et al. (2005) targeted HD209458 
(also with Spitzer) at 24 micron. The three flux measurements, showing 
relative contributions from 0.07\% to 0.28\%, are reasonably consistent with
the expectations from emission models, confirming a basic understanding of 
hot Jupiter atmospheric physics (Seager et al. 2005; Burrows, Hubeny 
\& Sudarsky 2005; Fortney et al. 2005). However, detailed conclusions can not be reached, and 
strong model degeneracies yet exist between e.g. the strength of molecular 
absorption, the Bond albedo, the redistribution of heat from the day to the 
night side of the planet, and the effective temperature of the planet. 

To allow a more detailed interpretation in terms of planetary atmosphere
physics and energy budgets, the flux measurements obtained so far ideally need 
to be complemented with observations at shorter wavelengths. In particular,
new measurements would be valuable from around the peak of the planets' 
energy distributions, between H$_2$O absorption bands at 1$-$4 $\mu$m. 
Although here the planet/star flux ratio is in general less favourable than 
at longer wavelengths, strong peaks in the spectral energy distribution of 
the planet around 2.2 and 3.9 $\mu$m are expected to 
significantly boost the signal, resulting in possible planet/star flux ratios 
at a 0.1\% level. This makes it potentially possible to observe the secondary
eclipse from the ground. So far, the highest accuracy from the ground has been
obtained using 'occultation spectroscopy' on HD209458b 
(Richardson et al. 2003a,b), 
aiming to detect the disappearance and reappearance of spectral features. 
This technique is sensitive only to the prominence of spectral 
features, not to the total flux of a planet. Their strongest limit has been
obtained on the contrast of the 2.2$\mu$m peak, at a level of 
$\sim$3$\times$$10^{-4}$ of the stellar flux.    

In this paper we present K-band photometry of the secondary eclipse of 
HD209458 obtained with UKIRT. 
 To achieve a 
photometric precision of 0.1\%, several challenges had to be overcome, 
in particular saturation of the K=6.3 star on the array, and 
non-linearity/aperture correction issues. The observation strategy
and data reduction are described in sections 2 \& 3, and the 
results and discussion are presented in sections 4 \& 5.

\section{Observations}

\begin{table}
\caption{Data on the exoplanet system HD209458, and two 
nearby stars used for flux calibration. The near-infrared data
are extracted from
the 2MASS point source catalogue. The data on HD209458b is taken from 
the extra-solar planet encyclopedia (http//www.obspm.fr/encycl/encycl.html).}
\begin{tabular}{l}\hline
HD209458 (2MASS22031077+1853036)\\
$\ \ \ \  ${\sl Stellar data}\\ 
$\ $Distance = 47 pc\\
$\ $Spectral type = G0V\\
$\ $Mass = 1.05 M$_{\rm{sun}}$\\
$\ $Radius = 1.12 R$_{\rm{sun}}$\\
$\ $Metallicity = 0.04 [Fe/H]\\
$\ $K$_{\rm{2MASS}}$=6.591\\
$\ $J$-$K$_{\rm{2MASS}}$=0.283\\
$\ \ \ \  $ {\sl Planet data}\\ 
$\ $Mass = 0.69$\pm$0.05 M$_{\rm{Jup}}$\\
$\ $Radius = 1.32$\pm$0.05 R$_{\rm{Jup}}$\\
$\ $Orbital Period =  3.52474541$\pm$0.00000025 days\\
$\ $T$_0$ (Mid-transit) = 2 452 854.825415$\pm$0.000060\\
$\ $Semi-major axis = 0.045 AU.\\\hline
REF. A (2MASS22030745+1851343)\\
$\ $K$_{\rm{2MASS}}$=9.912\\
$\ $J$-$K$_{\rm{2MASS}}$=0.310\\
\\
REF. B (2MASS22031559+1855064)\\
$\ $K$_{\rm{2MASS}}$=8.853\\
$\ $J$-$K$_{\rm{2MASS}}$=0.769\\ \hline
\end{tabular}
\end{table}

\begin{figure}
\psfig{figure=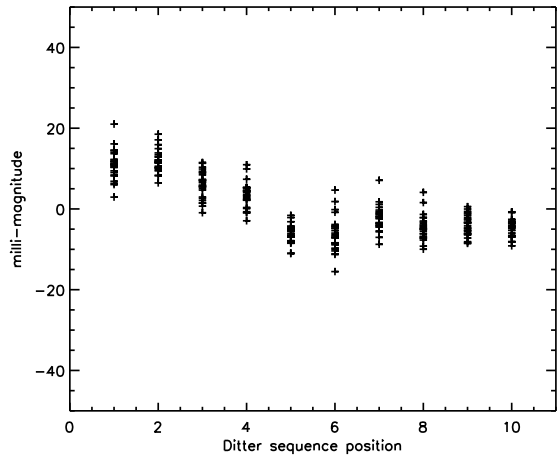, width=8cm}
\caption{\label{jitter} The relative K-band flux of HD209458b as function of 
the position in the jitter sequence for the observations obtained on 
September 13, 2004.}
\end{figure}

\begin{figure*}
\psfig{figure=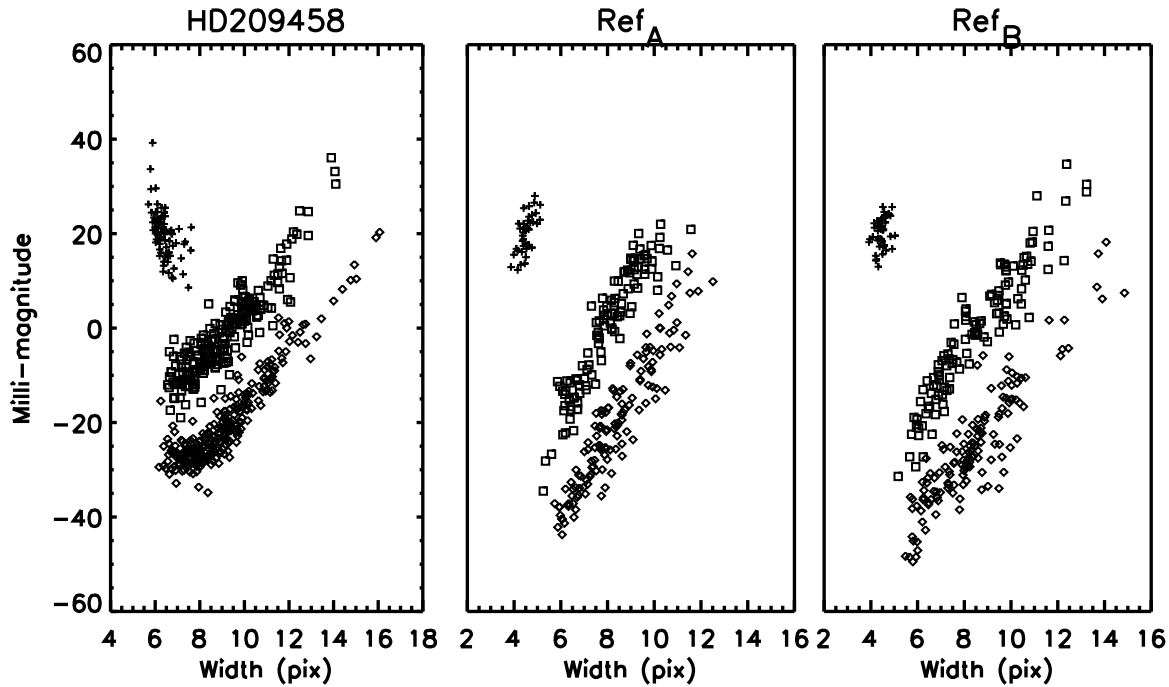, width=16cm}
\caption{\label{widths} The relative flux density of HD209458b (left), 
Ref$_{\rm{A}}$ (middle) and Ref$_{\rm{B}}$ (right) as function of the 
profile-width, for the July (crosses), August (squares), and September
(diamond) observations. The data-sets have been given separate offsets for
clarification. The dependencies are clearly different for the three 
sources, and most likely result from an interplay between residual 
non-linearity and aperture effects. }
\end{figure*}

We performed our K-band photometry on HD209458  with 
the 3.8m United Kingdom Infrared Telescope (UKIRT) on Mauna Kea, 
Hawaii, using UIST, the UKIRT 1-5 micron Imager Spectrometer.
HD209458b is of K=6.30 (see table 1), which means that under good seeing 
conditions it saturates the array within the shortest possible 
exposure time. It was therefore necessary to significantly defocus 
the telescope. UIST is equipped with a 1024x1024 InSb array, but only 
one quadrant was used to reduce readout times. In small pixel 
mode (0.06$''$/pixel) this results in a field of view of 30$\times$30 
arcseconds. Observing cycles of 15 minutes were defined in which subsequently 
the target, a reference star ref$_A$, again the target, and reference star 
ref$_B$ were observed. Each observation consists of a repeated 5-point jitter
(with $\pm$8$''$ offsets in RA and DEC), 
with the target and the reference stars falling onto the same area of the 
array. The reference stars  ref$_A$ and ref$_B$ are 2MASS22030745+1851343 
 and 2MASS22031559+1855064, at a 
distance 
of 101$''$ and 140$''$ to HD209458b respectively (see table 1 for details). 
The data were taken in NDSTARE mode, in which the array is reset, 
read immediately, and read again after the user-defined on-chip exposure time.
Exposure times  were 0.6, 6.0, and 3.0 seconds for the target, 
ref$_A$, and ref$_B$ respectively. 

Observations for this project were taken on three dates, all
 taken in service mode. 
Firstly, 1 hour test observations were performed on July 21, 2004 to
check the observing settings.
Subsequently, HD209458b was observed on August 23, from
06:06 UT to 10:18 UT, during which the eclipse occurred in the first half of 
the observation (assuming a circular orbit for HD209458b).
The dome had to be briefly closed due to local fog. 
 The target was observed again on the night of September 13,
from 07:22 UT to 11:27 UT. Here the eclipse occurred during 
the second part of the observations. A series of blank sky observations 
were used to construct a flat field at each epoch.

\section{Data reduction}

Data reduction and analysis was performed in IDL. 
Firstly, a non-linearity correction was applied to the data, using the 
predetermined relation,
\begin{equation}
g'=(1+8.26\times10^{-5}g-1.56\times10^{-8}g^2\\+1.173\times10^{-12}g^3)*g
\end{equation}
where $g$ and $g'$ are the measured and corrected data values 
(Sandy Leggett, privat communications).
A dark was subtracted from each individual frame,
after which it was divided through the flat 
field. The sky background was determined in the following way. 
First those areas on the array were identified that are sufficiently away
from the object at all jitter positions. The median 
background level in this area was determined for each frame, through which 
the frame was then normalised. These normalised frames were then used to 
determine any structure in the background by first masking out the object 
and then combining them. The resulting image was then multiplied by the 
median background level as determined above, to yield a sky-frame for 
each jitter point. This resulted in a dark subtracted, flat fielded,
and sky subtracted frame for each jitter point.

For each frame the total counts from the object was determined within 
a radius of 30, 40, 50, 60, and 70 pixels from its centre. In addition, 
a one-dimensional profile of the object was determined and fitted with 
a Lorentzian function, which was found to fit well. A first-order 
aperture correction was applied, as calculated from that part of the 
2-dimensional Lorentzian profile to fall outside the aperture, and the 
counts were converted to instrumental magnitudes. 
A small fraction (1-2\%) of the data points were not taken into account for 
further analysis, since their frames showed clear anomalies such as highly 
irregular light profiles indicating that the telescope underwent  
tracking errors during the exposure. 

\section{Results and analysis}

\begin{figure*}
\hbox{
\psfig{figure=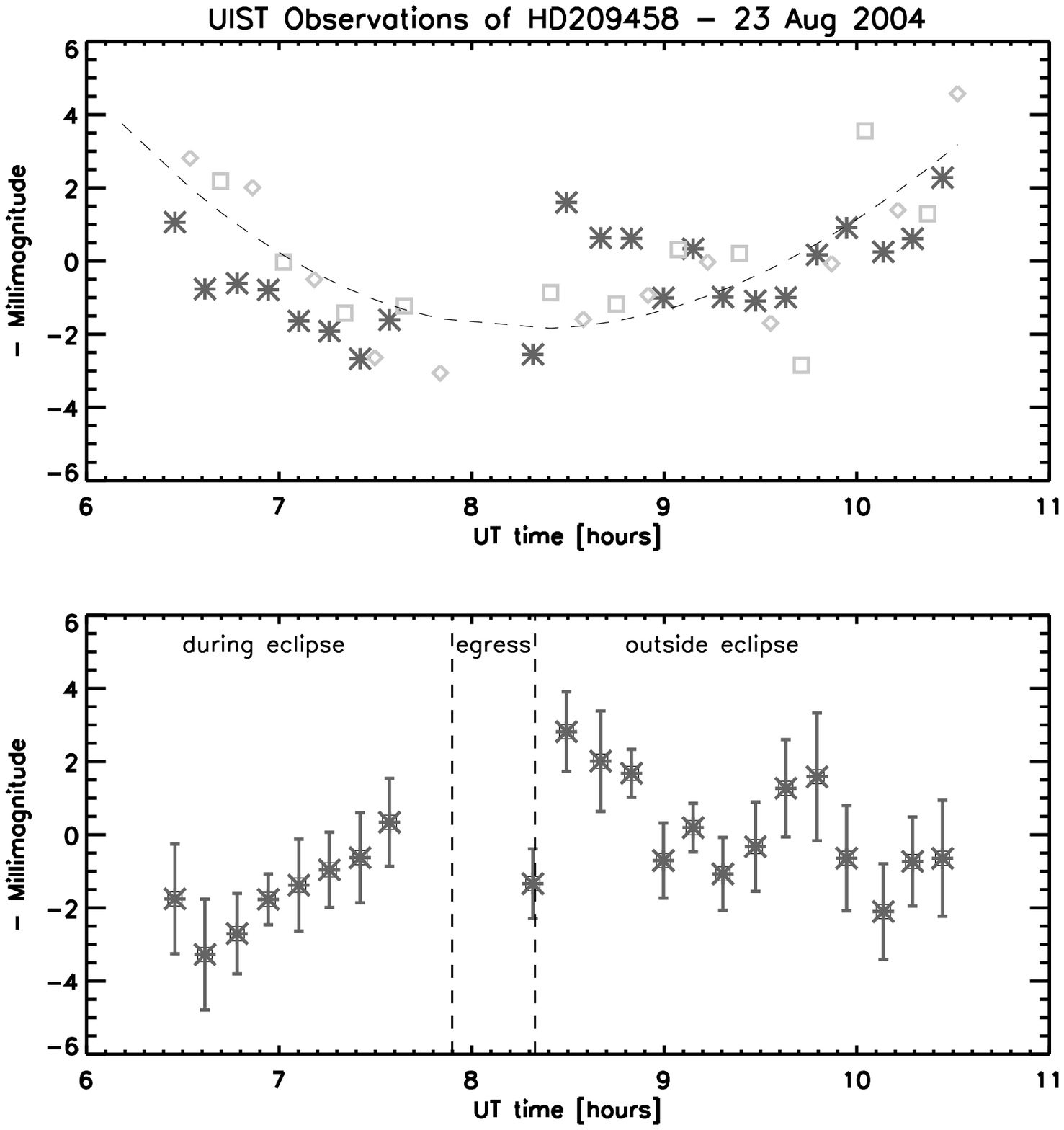, width=8cm}
\psfig{figure=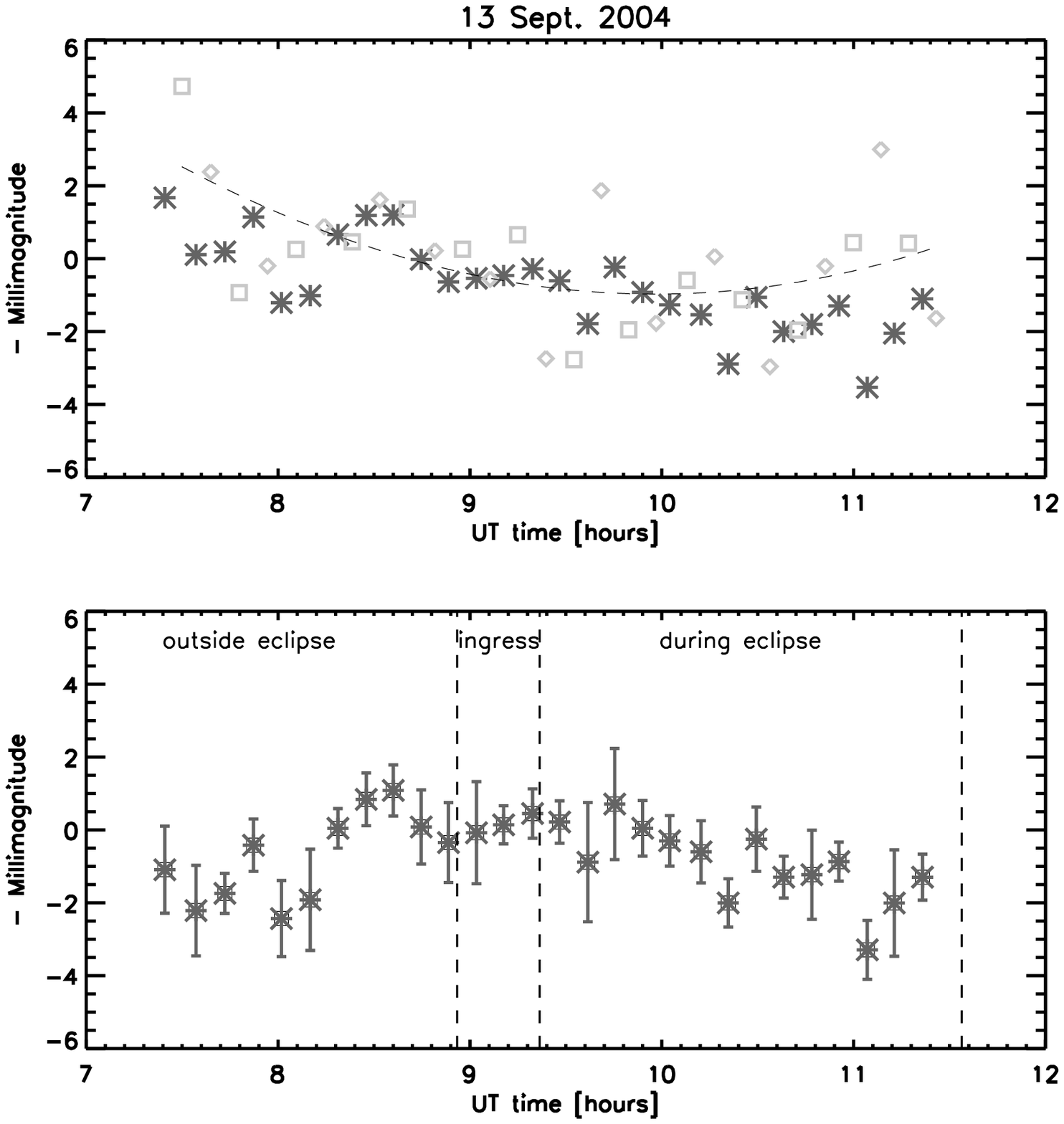, width=8cm}
}
\caption{\label{epochs} (left) The relative photometry obtained 
on August 23, 2004. The upper panel shows the data for the target (stars),
ref$_{\rm{A}}$ (squares), and ref$_{\rm{B}}$ (diamonds). The dashed line
shows the fit to the change in zero-point as function of time, as determined
from the two reference stars, which is used to determine the relative 
photometry for HD209458 as shown in the lower panel. The same data are 
shown for the September observations in the panels to the right.}
\end{figure*}

\subsubsection*{The test observations of July 2004}

The data taken on July 21 consists 4 cycles of target-ref$_A$-target-ref$_B$, 
resulting in 80 data frames for HD209458b, and 
40 frames for each of the reference stars. The data points, determined as 
described above, show a scatter of 22.6 milli-magnitude, but show a clear 
pattern (such as shown in Fig. \ref{jitter}). 
Firstly there is a clear dependence on jitter point, the first 
data point being on average 23 milli-magnitude fainter than the last data 
point in the jitter sequence. It is not clear what is causing this, 
but since it is found to be constant in time, it can be calibrated out. 
Secondly, there is a clear dependence of the magnitude of a 
data point on the width of the Lorentzian as fitted above. This dependence 
was found to be different for the target and the two reference stars 
(see the 'cross' symbols in Fig. \ref{widths}). The relation between
profile-width and magnitude for the target is such that the narrower the 
profile, the fainter the magnitude. This indicates that it may be caused by
a residual non-linearity effect, but residual aperture correction effect
may also play a role. The dependence between profile-width and magnitude goes
the other way around for ref$_A$ and ref$_B$. This different behaviour may 
be due to the fact that these objects 
were observed at lower count levels than the target, 
resulting in a different interplay between non-linearity and 
aperture correction effects. 
Note that the star profiles during the August and September observations 
(squares and diamonds in Fig. 2) have significantly broader widths, for 
which the width-magnitude relation of the target and reference stars do have
the same sign.
The width-magnitude relations for the 
target and the reference stars are fitted in an ad-hoc way. This results 
in a standard deviation of 5.3, 3.0, and 3.1 milli-magnitude for the 
target, ref$_A$ and ref$_B$ respectively.

\subsubsection*{The observations of August 2004}

Assuming that HD209458b has a circular orbit for which the secondary eclipse 
occurs exactly halfway between two transits, HD209458b  was 
mid-eclipse on August 23, 2004 at 6:47 UT (taking into account light travel 
time in our solar system). Twelve cycles of 
target-ref$_A$-target-ref$_B$ were observed on that night from 
06:06 UT to 10:18 UT. A similar pattern in magnitude as function of 
jitter position as found in July, was removed from the data. 
A new complication compared to the 
the test observations was a significant change in airmass over the four hours
of observations. Therefore the flux zero-point changed slowly causing a 
difference
of 31 milli-magnitude between the beginning and the end of the observations. 
An airmass correction was determined for ref$_A$ and applied to all three
objects. The star profiles were found to vary more wildly between frame 
to frame than during the test observations, with the widths varying  
from 6 to 15 pixels. Since the variation was  quasi-random with time, 
the width-magnitude relations for the 
target and the reference stars could be fitted and used to correct their 
magnitude in a similar way as for the test observations (but see below). 
When the corrections for airmass and profile width are applied, they result in 
an improved correction for jitter position, which in its way results in a 
better correction as function of airmass and profile width $-$ and so forth.
This iterative process quickly results in optimal solutions for all three
relations, resulting in a dispersion of 3.6, 4.1, and 4.6 milli-magnitude for 
the target, ref$_A$ and ref$_B$ respectively. 
The 10 data points of each jitter sequence of each source are averaged to 
increase the signal-to-noise. The resulting photometry is shown in figure 
\ref{epochs},
with the lower panel showing the target data points relative to those of the 
two reference stars. For a circular orbit it is expected that the planet would 
reappear from behind its host star between 7:46 and 8:10 UT. The data points 
of HD209458 before this time are on average 1.4$\pm$2.0 millimagnitude fainter 
($-$0.13$\pm$0.18\%) than those after. 

\subsubsection*{The observations of September 2004}

On September 13, HD209458b underwent a secondary eclipse centered at 
10:13 UT. Fourteen cycles of target-ref$_A$-target-ref$_B$ were observed on 
that night from 07:22 UT to 11:27 UT.
A similar pattern in magnitude as function of jitter position was 
found as for the two epochs before, and removed from the data.
While these observations were more ideally timed, with the objects 
located at low airmass throughout the observing session, the seeing 
dramatically improved during the four hours of observing, causing the 
profile-widths to change from on average of 14.5 pixels in the third cycle, 
to 7.0 pixels in the last cycle. This resulted in a degeneracy between
a possible flux density change from the first to the second half of the data 
set for HD209458 (as expected for an eclipse), and the precise relation 
between seeing and measured flux density. Indeed, if a fake signal is 
introduced that mimics 
an eclipse, the reduction procedure as described above subsequently 
removes most of it. This problem was solved by fitting the width-magnitude
relation for the in- and out-eclipse data separately, allowing a flux
offset between the two relations but keeping the flux-dependence on 
width the same for the two sub-sets, hence adding one extra variable. 
In this way the reduction procedure leaves a fake signal intact. 
Note that analysing the August dataset in this way does not change the result 
obtained above. The in-eclipse flux points are found to be on average
1.1$\pm$1.1 milli-mag fainter ($-$0.10$\pm$0.10\%) than outside the 
eclipse (see Fig. 3).

\section{Discussion and conclusions}

\begin{figure}
\psfig{figure=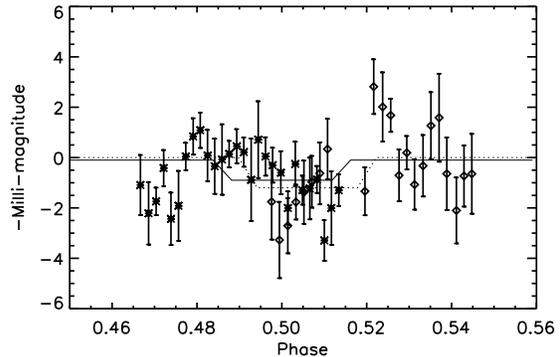, width=8cm}
\caption{\label{final} The combined dataset for the August 23 (diamonds) 
and September 13 (stars) observations. Note that no flux offset was applied 
between the two datasets. 
The solid line indicates the best fitting eclipse for a fixed circular orbit,
and the dotted line while keeping the eclipse timing as a free parameter. 
The combined flux decrement is not 
statistically significant due to the uncertainty in the flux-level outside
the eclipse, and the uncertainty in the flux offset between the two datasets.}
\end{figure}

Our observations of the secondary eclipse of HD209458b 
have resulted in a photometric accuracy of $\sim$0.12\% per 15 min. cycle 
compared
to two nearby reference stars. No firm detections of the eclipses 
were obtained, with the first observation showing a flux decrement
of $-$0.13$\pm$0.18\% during the time of the eclipse, and the 
second of $-$0.10$\pm$0.10\%. In Fig. \ref{final} we show the 
combined dataset. Note that the average brightness-difference  
between the two reference stars and HD209458 changed by only $\sim$0.3 
milli-magnitude between the two epochs and was not corrected for. 
An eclipse-profile was fitted to this combined dataset, for 
a circular orbit (solid line), and for a non-circular orbit (dotted line),
allowing the precise timing of the eclipse to change. This because a
non-zero orbital eccentricity can produce a shift in the separation 
of the times of transit and eclipse away from half-period, although 
Deming et al (2005) find that the secondary eclipse of HD209458 occurs 
at the mid-point between transits to within $\pm$7 minutes. Neither the 
fit for a fixed circular orbit, nor the fit for a free eclipse time give 
a significantly better fit than obtained for the individual data sets. 
By eye, one may be inclined to see a hint of an eclipse, but the 
uncertainty in the average flux level outside the eclipse (combined with
the uncertainty in the flux offset between the two datasets) makes this 
signal not statistically significant. 

\begin{figure}
\psfig{figure=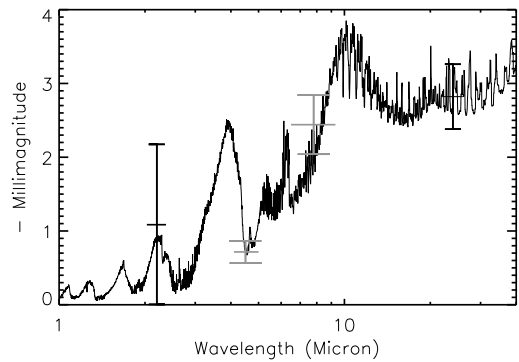, width=8cm}
\caption{\label{model} Eclipse depth as function of wavelength for 
HD209458, assuming the baseline model-atmosphere as determined by
Sudarsky et al. (2003), but scaled to the eclipse-depth measurement
of Deming et al. (2005) at 24 $\mu$m. The two grey data points indicate
the measurements from Charbonneau et al. (2005) of the exoplanet system
TrES-1 (which flux ratios are not expected to differ much from that of 
HD209458).}
\end{figure}

Figure \ref{model} shows a model of the eclipse depth as function of 
wavelengths for
HD209458b, as calculated by Sudarsky, Burrows, \& Hubeny (2003), scaled to 
fit the 24 $\mu$m measurement by Deming et al. (2005). Over-plotted 
are also the two eclipse-depth measurements of Charbonneau et al. (2005) for
TrES-1, which is expected to show very similar eclipses.
It shows that for this baseline model, at 2.2 $\mu$m an eclipse-depth of 
$\sim$1 milli-magnitude is expected, while the observations described in this 
paper provide an upper limit of $\sim$2.1 milli-magnitude.
Note that Richardson et al. (2003), who used occultation spectroscopy to
measure the strength of the feature at 2.2 $\mu$m, claim to detect no such 
peak at a level of $\sim$3$\times$10$^{-4}$ of the stellar flux, rejecting the 
Sudarsky et al. baseline model. 

\subsection*{Future prospects}

We believe there are several ways to further improve the 
relative photometric precision of $\sim$1 milli-magnitude as obtained
here. First of all, HD209458 is not anymore the most ideal candidate for
ground-based photometry. \footnote{The referee, Drake Deming, instead suggests
to monitor HD209458 with a smaller aperture telescope, avoiding saturation
and allowing to simultaneously observe bright comparison stars in its 
wider field of view.} TrES-1 (discovered in August 2004; Alonso et al.
2004) is of K=9.8, with two nearby stars with similar magnitudes and colours.
This means that the telescope does not have to be defocused and that 
the aperture correction problems may be less prominent. 
Furthermore, for TrES-1, 
the nearby stars mean that the calibrators can be observed in the same
frame as the target, which we believe will improve the photometry considerably.
Frame-to-frame variations in the observations for HD209458b indicate
that, if a prefect calibration of the throughput of the telescope+atmosphere
and the sensitivity of the array had been possible, a further improvement 
of 30-40\% in precision would be possible. Furthermore, observing the complete
eclipse, plus a significant period before and after the dis- and reappearance 
of the planet would allow possible slowly varying drifts in flux density,
e.g. due to intrinsic variability of calibrations stars.

In addition to secondary eclipse observations, high precision near-infrared
photometry has also a potentially interesting application in 
exoplanet transit surveys, in particular in the light of the recent 
advent of large infrared arrays such as WFCAM on UKIRT, WIRCam on 
the CFHT, and in the near future, VISTA.  We show that relative photometry 
in K-band is possible down to a milli-magnitude level. Near-infrared transit 
surveys are particularly interesting for probing planets around 
late M and L dwarfs, which are significantly brighter in K-band than 
in the optical. Their relatively small size will make any planet transit 
more pronounced than for solar type stars. For example, a transit of an earth 
size planet around an M8 dwarf star (R$\sim$0.2R$_{\rm{sun}}$), would result
in a transit depth of $\sim$0.2\%. Hence, it may potentially be possible to 
detect earth-type planets with near-infrared transit surveys.

\section*{acknowledgements}

The author wishes to thank the UKIRT staff, in particular 
Andy Adamson and Sandy Leggett, for their help and suggestions 
during the project. Also thanks to Katherine Inskip for doing 
the September observations in service.  
The United Kingdom Infrared Telescope is operated by the Joint 
Astronomy Centre on behalf of the U.K. Particle Physics and 
Astronomy Research Council.

\label{lastpage}


\begin{thebibliography}{99}

\bibitem{} Burrows A., Hubeny I., \& Sudarsky D., 2005, ApJ 625, L135
\bibitem{} Charbonneau D., Brown T.M., Noyes R.W., Gilligand R.L., 2002,
            ApJ 568, 377
\bibitem{} Charbonneau D., Allen L., Megeath S., Torres G., Alonso R., 
           Brown T., Gilliland R., Latham D., Mandushev G., O'Donovan F., 
           Sozzetti A., 2005, ApJ, accepted
\bibitem{} Deming D., Seager S., Richardson J. \& Harrington J.  
           2005, Nature , 434 , 740
\bibitem{} Fortney J.J., Marley M.S., Lodders K., Saumon D., Freedman R.,
           2005, ApJ 627, L69
 \bibitem{} Marcy G., Butler R.P., Fischer D., Vogt S., Wright J.T., Tinney 
C.G., Jones H.R.A, 2005, Progress of Theoretical Physics Supplement, in press
\bibitem{} Richardson, L.~J., Deming, D., Wiedemann, G., Goukenleuque, C., 
           Steyert, D., Harrington, J., \& Esposito, L.~W., 2003a, 
           ApJ, 584, 1053 
\bibitem{} Richardson J.L., Deming D., Seager S., 2003b, ApJ 597, 581
\bibitem{} Seager S., Richardson L.J., Hansen B.M.S., Menou K., Cho J.Y-K.,
 Deming D., 2005, submitted to ApJ
\bibitem{} Sudarsky D., Burrows A., \& Hubeny I., 2003, ApJ 588, 1121
\bibitem{} Vidal-Madjar A, Lecavelier des Etangs A., D\'{e}sert J.-M., 
            Ballester G.E., Ferlet R., H\'{e}brard G., Mayor M., Nature 422,
            143
\bibitem{} Vidal-Madjar A, D\'{e}sert J.-M., Lecavelier des Etangs A.,
            H\'{e}brard G., Ballester G.E., Ehrenreich D., Ferlet R.,
            McConnell J.C., Mayor M., Parkinson C.D., 2004, ApJL submitted
\end{thebibliography}
\end{document}